\documentclass[letterpaper,journal]{IEEEtran}
\IEEEoverridecommandlockouts
\usepackage{amsmath,amsfonts,amssymb}
\usepackage{siunitx}
\usepackage[noend]{algpseudocode}
\usepackage{algorithm}
\usepackage{array}
\usepackage[caption=false,font=footnotesize,labelfont=sf,textfont=sf]{subfig}
\usepackage{textcomp}
\usepackage{stfloats}
\usepackage[hyphens]{url}
\usepackage{verbatim}
\usepackage{graphicx}
\graphicspath{{images/}}
\usepackage{cite}
\usepackage{hyperref}
\usepackage{tikz}
\usetikzlibrary{arrows}
\usetikzlibrary{shapes.multipart}
\usetikzlibrary{plotmarks}
\usetikzlibrary{patterns}
\RequirePackage{xcolor}

\newcommand{\CP}{\mathrm{P}}
\newcommand{\vv}{\mathbf{v}}
\newcommand{\ww}{\mathbf{w}}
\newcommand{\Z}{\mathrm{Z}}
\newcommand{\Srm}{\mathrm{S}}
\newcommand{\F}{\mathbf{F}}
\newcommand{\f}{\mathbf{f}}
\newcommand{\Tb}{\mathbf{T}}

\IEEEoverridecommandlockouts
\newcommand\copyrighttext{%
\footnotesize \textcopyright \enspace 2024 IEEE. Personal use of this material is permitted. Permission from IEEE must be obtained for all other uses, in any current or future media, including reprinting/republishing this material for advertising or promotional purposes, creating new collective works, for resale or redistribution to servers or lists, or reuse of any copyrighted component of this work in other works. DOI: \href{https://doi.org/10.1109/LCOMM.2024.3397489}{10.1109/LCOMM.2024.3397489}
}
\newcommand\copyrightnotice{%
\begin{tikzpicture}[remember picture,overlay]
\node[anchor=south] at (current page.south) {\fbox{\parbox{\dimexpr\textwidth-\fboxsep-\fboxrule\relax}{\copyrighttext}}};
\end{tikzpicture}%
}

\begin{document}

\title{A Scheduler for Real-Time Service in Wi-Fi 8 Multi-AP Networks with Parameterized Spatial Reuse}

\author{Kirill Chemrov, Dmitry Bankov, Andrey Lyakhov, Evgeny Khorov
\thanks{Kirill Chemrov, Dmitry Bankov, Andrey Lyakhov, and Evgeny Khorov are with the Institute for Information Transmission Problems of the Russian Academy of Sciences, Moscow 127051, Russia (emails: lastname@wireless.iitp.ru).}
}

\maketitle
\copyrightnotice

\begin{abstract}
	Real-time applications (RTAs) require low delays and impose a significant challenge to Wi-Fi.
	In Wi-Fi, high delays are often caused by waiting for the channel to become idle.
	This problem can be solved with Parameterized Spatial Reuse (PSR), which allows a station (STA) to transmit its frame with reduced power simultaneously with a triggered uplink transmission in an overlapping network.
	The PSR opportunity depends on the pathloss between involved STAs, so the same transmission may allow PSR for one STA but not for another one.
	Thus, to satisfy tight delay constraints in dense overlapping networks, access points (APs) in the same area shall often allow PSR for every STA with RTA traffic.
	This letter proposes a fast scheduler enabling frequent PSR transmissions for RTA traffic. The scheduler uses Multi-AP coordination, the feature of upcoming Wi-Fi~8.
	With simulations, we show that it almost halves the delay for RTA traffic and does not deteriorate the quality of service for other traffic compared with an airtime fairness scheduler.
\end{abstract}

\begin{IEEEkeywords}
	Wi-Fi 8, Real-Time Applications, Channel Resource Scheduling, Parameterized Spatial Reuse, Multi-AP
\end{IEEEkeywords}

\section{Introduction}
\label{sec:intro}
The traffic of real-time applications (RTAs), such as virtual and augmented reality, remote control, industrial automation, cloud gaming, etc, requires low delays and high reliability. For example, Wi-Fi beyond 2030 shall support immersive communications with 99.9\% of packets being served with delays below \SI{20}{ms}~\cite{giordano2023wifi8}. These requirements are challenging for modern Wi-Fi. 
First, Wi-Fi uses unlicensed spectrum, suffering from interference, which induces packet losses.
Second, Wi-Fi uses random channel access, which results in high delays caused by waiting for the end of ongoing transmissions and collisions~\cite{chemrov2022smart}.
Third, the density of Wi-Fi networks (or Basic Service Sets, BSSs) is growing rapidly, leading to many overlapping BSSs, and inducing more interference.

Recently launched 802.11bn (known as Wi-Fi~8~\cite{giordano2023wifi8}), considers RTA as a key use case and studies how to improve the worst-case latency for RTA traffic.
A promising way is extending several features introduced in Wi-Fi~6~\cite{khorov2018tutorial} with multi-AP coordination, a killer feature of Wi-Fi~8.

The first feature of Wi-Fi 6 is trigger-based (TB) transmission.
This mechanism allows an Access Point (AP) to send a trigger frame (TF) and allocate channel resources for uplink transmissions of some stations (STAs).
TB transmissions help to avoid contention between client STAs of the same BSS.

The second feature is Parameterized Spatial Reuse (PSR), which improves the performance of overlapping BSSs~\cite{rodrigues2020latency, wilhelmi2021spatial}.
PSR allows a STA to transmit during an ongoing TB transmission from another BSS.
For that, the AP initiates a TB transmission with a TF, notifying the surrounding STAs about the PSR opportunity.
If a STA wants to use the PSR opportunity, it reduces its transmit power according to information in the TF to avoid high interference with the TB transmission.
PSR can be extremely useful for RTA because it significantly accelerates the access to the channel when it is busy~\cite{rodrigues2020latency}.

PSR should only be used if the power restriction is not too strong and the delivery is reliable.
However, this restriction depends on the pathloss between STAs, so not all TB transmissions are favorable for the PSR transmission of a particular STA.
As a result, the efficiency of PSR in reducing delays depends on the schedule of TB transmissions.
Therefore, APs need to use multi-AP coordination for information exchange to adjust the schedule, taking into account the PSR opportunity for STAs of overlapping BSSs with RTA traffic.

Note that although existing works, e.g.,~\cite{wilhelmi2021spatial, chemrov2023support}, show that PSR is favorable for RTA, they do not provide an algorithm to arrange transmissions so that the requirements of RTA STAs are satisfied.
Moreover, the approach from~\cite{chemrov2023support} implies simple alternation of TB transmissions that allow and disallow PSR, which leads to unfairness in resource allocation.
For example, if too many STAs allow PSR, then each of them receives more channel time than each of those that disallow PSR.
In this letter, we design a new scheduling approach that, in comparison with the airtime fairness scheduler, reduces delays for several RTA STAs without degradation of throughput and fairness for STAs with delay-insensitive traffic.

The rest of the letter is structured as follows.
In Section~\ref{sec:problem}, we describe the scenario and state the problem.
Section~\ref{sec:scheduler} contains the proposed scheduling approach with the optimization problem statement and its greedy solution. 
In Section~\ref{sec:results}, we discuss numerical results.
Section~\ref{sec:conclusion} concludes the letter.

\section{Scenario and Problem Statement}
\label{sec:problem}

Consider two overlapping BSSs sharing the same channel.

The first BSS (hereinafter referred to as \emph{non-RTA BSS}) includes $N$ non-RTA STAs and a non-RTA AP, each of which generates non-RTA broadband traffic.
The AP controls uplink transmissions with TF to prevent collisions: after a portion of downlink data, it transmits a TF to start an uplink transmission. 

The second one (hereinafter referred to as \emph{RTA BSS}) consists of an RTA AP and $M$ RTA STAs, which generate uplink RTA traffic.
As the AP can access the channel with high priority,  serving RTA downlink traffic is not as challenging as serving uplink RTA one.
Thus, we consider only uplink traffic in the RTA BSS.
As the RTA AP does not know about the presence of RTA data at STAs, it does not use TB transmissions.

The non-RTA AP provides the PSR opportunity during some TB uplink transmissions.
Note that the PSR opportunity cannot be provided during downlink transmissions.
The PSR opportunity may set too stringent power restrictions for some RTA STAs, which makes successful delivery impossible.
Let us call a non-RTA STA PSR-favorable for a particular RTA STA if, during its TB transmissions, PSR is allowed with such power restriction that allows signal-to-interference-plus-noise ratio (SINR) of the RTA STA's transmission to exceed some threshold $\text{SINR}_{th}$.
Otherwise, if a non-RTA STA disallows PSR or allows only unreliable PSR transmission for an RTA STA, we call it PSR-unfavorable for this RTA STA.

Without coordination between APs, the non-RTA AP is unaware of RTA STAs in the overlapping BSS and does not know which non-RTA STAs are PSR-favorable for them.
So, the non-RTA AP performs some scheduler, e.g., the airtime fairness one, which on average allocates equal channel time for each STA~\cite{joshi2008airtime}.
As a result, the AP may schedule STAs that are PSR-unfavorable for some RTA STAs several times in a row, which increases worst-case channel access delays.

Thanks to multi-AP coordination considered for Wi-Fi~8, we can design better scheduling algorithms.
Specifically, APs perform the following PSR-aware classification for each RTA STA.
The non-RTA AP communicates with associated STAs using TB transmissions.
During this data exchange, the non-RTA AP provides the RTA AP with an allowed interference level using TF for each STA that allows PSR.
Note that the allowed interference level can vary per non-RTA STA as it depends on the used modulation and coding scheme (MCS). 
At the same time, each RTA STA measures the received power level of the TF from non-RTA AP and reports it to the RTA AP, aggregating this information with data.
Then, the RTA AP measures the received power level of TB transmission from each non-RTA STA and calculates the expected SINR of considered RTA STA transmission with the PSR opportunity granted during each non-RTA STA TB transmission.
Note that the procedure does not require additional regular signaling, as it is enough to measure the received power of data transmissions and TF.
A non-RTA STA is considered PSR-favorable if it meets the following condition: $\text{SINR} > \text{SINR}_{th}$.

This procedure is repeated several times to make the classification more robust in mobile scenarios, e.g., when a VR player moves. 
A non-RTA STA is considered PSR-favorable only if the PSR criterion is met for all recent measurements.
Finally, the RTA AP reports to the non-RTA AP the set of PSR-favorable STAs, via wireless or wired backhaul.  While wireless backhaul is under development~\cite{sigurd2022cofdma}, some multi-AP solutions already use the wired one~\cite{zehl2016resfi}. 
In any case, the procedure does not induce much control traffic, as only significant changes in SINR affecting the classification generate new reports.

We state a problem to develop a scheduler for non-RTA BSS that uses multi-AP coordination to improve the quality of service (QoS) for RTA BSS without QoS degradation for non-RTA BSS.
As a QoS metric for RTA traffic, we consider the $Q$ (0.999) quantile of delay among all STAs, where the delay is the time between a frame enqueuing and its delivery.
For non-RTA traffic, we consider the average throughput and Jain's fairness index~\cite{jain1984quantitative} of throughput: the ratio of the squared average throughput to the average of the squared throughput.

\section{Proposed Scheduling Approach}
\label{sec:scheduler}

To solve the stated problem, we design a scheduler that determines the order of uplink non-RTA transmissions.
To guarantee the fairness of resource allocation for non-RTA STAs, this order is repeated but in every repetition, each non-RTA STA transmits only once.
Among all possible permutations of non-RTA STAs, the scheduler minimizes the worst-case distance between PSR-favorable transmissions for each RTA STA, which reduces the Q-quantile of RTA delays.
This approach makes the scheduler independent of the type of RTA traffic, e.g., whether it is sporadic or regular.
Moreover, since the scheduler only reorders the non-RTA STA transmissions, it also guarantees that the average delay does not notably degrade for non-RTA STAs.
In Section~\ref{ssec:math}, we state an optimization problem to find the order of non-RTA STAs.
Then, in Section~\ref{ssec:greedy}, we design a greedy solution for it.

\subsection{System Analysis}
\label{ssec:math}
Let us formalize the problem stated in Section~\ref{sec:problem}.
To characterize a non-RTA STA $i$, we use a favorability vector $\f_i =\{f_{ij}\}_{j=1}^M$, $i\in 1..N$. 
We set $f_{ij} = 1$, if a non-RTA STA $i$ is PSR-favorable for RTA STA $j$, otherwise, we set $f_{ij} = 0$.
Let the scheduler make an order of active STAs, i.e., STAs with traffic, $\{i_1, i_2, ..., i_N\}$.
To describe the scheduling of non-RTA BSS transmissions for one period (column order) and the corresponding PSR opportunities for RTA STAs, we use vectors $\f_i$ in this order to build a matrix: $\F = \left[\f_{i_1}|\f_{i_2}|...|\f_{i_N}\right]$, where $[\mathbf{A}|\mathbf{B}]$ denotes concatenation of matrices $\mathbf{A}$ and $\mathbf{B}$.

Let $\Z_i(\F)$ be the maximum number of consecutive zeros in the $i$-th row of the periodically repeated matrix $[...|\F|\F|\F|...]$.
For example, for a row $(0, 1, 0, 1, 0, 0)$,  $\Z_i =3$.
Note that it is enough to only consider $[\F|\F]$ instead of $[...|\F|\F|\F|...]$.
If the row $i$ contains $k$ consecutive zeros, the RTA STA $i$ cannot make a PSR transmission for the time of $k$ non-RTA transmissions, which results in delays as long as the duration of $k$ non-RTA transmissions.
Thus, to minimize the worst-case channel access delays for the RTA STA $i$, we minimize $\Z_i(\F)$ over all column permutations of matrix $\left[\f_{1}|\f_{2}|...|\f_{N}\right]$. 
We denote these all permutations as  $\CP\{\f_i\}$.

As it is not always possible to find a permutation that is optimal for all RTA STAs, we state a problem to minimize $\max_{i\in1..M}\Z_i(\F)$ over all $\F \in \CP\{\f_i\}$.
If several permutations give the same $\max_{i\in1..M}\Z_i(\F)$, we minimize the next maximum, etc. 
Let vector $\Srm(\F)$ contain $\{\Z_i(\F)\}_{i=1}^{M}$ in descending order.
We state the lexicographic optimization problem:
\begin{equation}\label{eq:opt_problem}
	\Srm(\F) \rightarrow \underset{\F\in \CP\{\f_i\}}{\mathrm{lex~min}}
\end{equation}

Note that the rows in $\F$ that consist of all zeros or all ones do not change with column permutation.
They do not affect the solution of the optimization problem and thus can be omitted while solving it.
We further assume that all the $M$ rows in $\F$ have at least one $0$ and one $1$.
The optimization problem \eqref{eq:opt_problem} can be solved with a brute force approach,
but its complexity is $\mathcal{O}(M\times N!)$, so we propose a fast greedy algorithm.

\subsection{Greedy Scheduler}
\label{ssec:greedy}
\begin{algorithm}[tb]
	\caption{Greedy Scheduler}\label{alg:greedy}
	\begin{algorithmic}[1]
		\Require{favorability vectors $\{\f_{i}\}_{i=1}^{N}$}
		\Ensure{greedy optimal order of vectors $\F^*$}
		\State $\F^* \gets [\f_1|\f_2]$ \label{line:init}
		\For{$i$ in $3..N$} \label{line:fori}
		\State $\Tb^* \gets \F^* $ \label{line:Tinit1}
		\State insert $\f_i$ in $\Tb^*$ after column $1$ \label{line:Tinit2}
		\For{$j$ in $2..(i-1)$}  \label{line:forj}
		\State $\Tb \gets \F^*$  \label{line:Finit1}
		\State insert $\f_i$ in $\Tb$ after column $j$ \label{line:Finit2}
		\If{$\textsc{lexigraphicallyless}\left(\Srm(\Tb),~\Srm(\Tb^*)\right)$} \label{line:iflexless1}
		\State $\Tb^* \gets \Tb$
		\EndIf	\label{line:iflexless2}
		\EndFor \label{line:Fsetend}
		\State $\F^* \gets \Tb^*$ \label{line:TtoF}
		\EndFor \label{line:foriend}
		\item[] \vspace{-0.75\baselineskip}
		\State \textbf{Input:} vectors $\vv=\{v_{m}\}_{m=1}^{M}$ and $\ww=\{w_{m}\}_{m=1}^{M}$ \label{line:lexless1}
		\State \textbf{Output:}	whether $\vv$ is lexicographically less than $\ww$
		\Function{lexicographicallyless}{$\vv$, $\ww$} 
		\For{$m$ in $1..M$}
		\If{$v_m < w_m$}
		\State \Return true
		\ElsIf{$v_m > w_m$}
		\State \Return false
		\EndIf
		\EndFor
		\State \Return false
		\EndFunction	\label{line:lexless2}
		
	\end{algorithmic}
\end{algorithm}

The idea of the greedy algorithm is to take vectors one by one and insert them in such a place of the result matrix that minimizes the objective function $\Srm(...)$ for the current matrix, see Algorithm~\ref{alg:greedy}.
The algorithm takes as input a set of favorability vectors of active STAs $\{\f_{i}\}_{i=1}^{N}$ and outputs the greedy-optimal sequence of vectors concatenated in matrix $\F^*$.

We initialize the resulting matrix $\F^*$ by concatenating two first vectors (line~\ref{line:init}).
For each remaining vector $\f_i$, we find the best position in the matrix $\F^*$ (lines~\ref{line:fori}--\ref{line:foriend}).

Let matrix $\Tb^*$ store the current optimal position of vector $\f_i$. We initialize $\Tb^*$ with the matrix $\F^*$ and insert the vector $\f_i $ after column~1~(lines~\ref{line:Tinit1}--\ref{line:Tinit2}).
Then, we iterate through the remaining possible positions $j$ of the vector $\f_i$ (line~\ref{line:forj}).
Variable $\Tb$ stores the copy of the matrix $\F^*$ with the vector $\f_i$ inserted after column $j$ (lines~\ref{line:Finit1}--\ref{line:Finit2}).
If the objective function $\Srm(\Tb)$ is lexicographically less then $\Tb^*$, i.e., we find a better position for $\f_i$, we overwrite $\Tb^*$ with $\Tb$ (lines~\ref{line:iflexless1}--\ref{line:iflexless2}).
	
When all the positions are exhausted, $\Tb^*$ stores the matrix with the best position of $\f_i$.
So, we overwrite the resulting matrix $\F^*$ with $\Tb^*$ and consider the next vector $\f_{i+1}$ (line~\ref{line:TtoF}).

When all the vectors from $\{\f_{i}\}_{i=1}^{N}$ are exhausted, we get the final greedy-optimal matrix $\F^*$,
which column order determines the scheduling order of non-RTA STAs.

Note that the vectors $\{\f_{i}\}_{i=1}^{N}$ can be added to the matrix $\F^*$ in an arbitrary order.
We have tried several way to sort  $\{\f_{i}\}_{i=1}^{N}$, e.g.,  by their Hamming weights.
However, the numerical results for all of them turned out to be indistinguishable from the case without the preliminary sorting considered in Fig.~\ref{fig:plot}.

The complexity of the lexicographical comparison of vectors of size $M$ is $\mathcal{O}(M)$.
As it is performed in a double nested loop, the final complexity of the greedy algorithm is $\mathcal{O}(M\times N^2)$, which is significantly better than brute force.

Although we consider only two overlapping BSSs to simplify the explanation, the developed scheduler is valid for several overlapping BSSs. In this case, it creates a joint schedule for these BSSs, which can be implemented with multi-AP coordination, e.g., using the extended TXOP sharing considered for Wi-Fi~8 \cite{liwen2023txs}.
A controller that manages the joint schedule collects the favorability vectors of STAs in all BSSs and runs the scheduler for all the received vectors. 
Thus, adding more BSSs only increases $N$ in the above algorithm.

\section{Results}
\label{sec:results}

\begin{figure}[tb]
	\centering
	\includegraphics[width=0.9\linewidth]{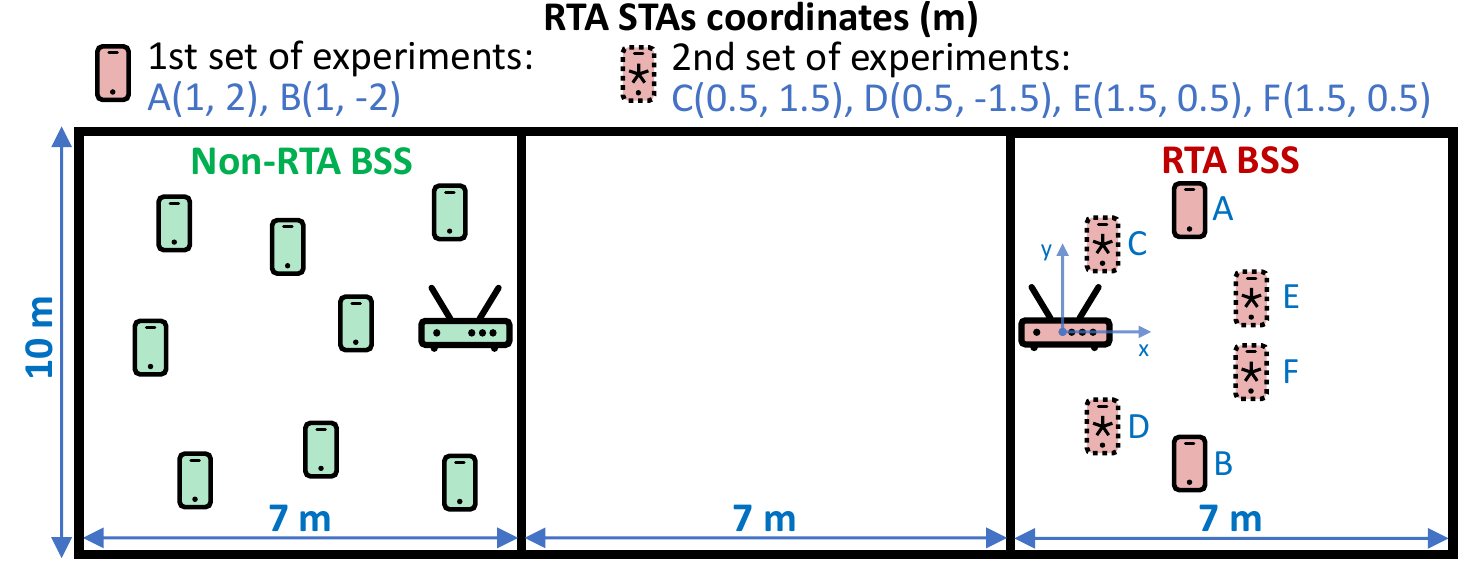}
	\caption{\label{fig:scenario} Scheme of the scenario.}
\end{figure}

We study the efficiency of the proposed scheduler with extensive experiments in ns-3~\cite{nsnam}. 
Consider an apartment building, and let the Wi-Fi networks of two apartments on the same floor use the same band.
Specifically, devices of two overlapping BSSs are located in two apartments seven meters from each other, see Fig.~\ref{fig:scenario}, which is reasonable as neighboring apartments likely use different channels.
The non-RTA AP is near the right wall, and positions of $N=8$ non-RTA STAs are randomly scattered inside the $10 \times 7$ meters apartment.
Non-RTA devices have saturated traffic of packets with the size of 1000 bytes, which are aggregated and transmitted with MCS~8.
In fact, not all the STAs can have saturated traffic, so we consider only those STAs that have data to transmit.
The non-RTA AP gains a transmission opportunity of $\approx \SI{5}{ms}$ and uses it equally for downlink and uplink transmissions.

\begin{table}[t]
	\caption{Modeling Parameters}\label{tab:parameters}
	\begin{center}
		\begin{tabular}{|l|l|}
			\hline
			\textbf{Parameter} & \textbf{Description} \\
			\hline
			Channel width & \SI{20}{MHz} \\
			Number of spatial streams & 1 \\
			Baseline scheduler & Airtime fairness \\
			Transmit power & AP: \SI{20}{dBm}, STA: \SI{15}{dBm} \\
			PSR-favorable threshold $\text{SINR}_{th}$ & \SI{3}{dB} \\
			PSR safety margin & \SI{1}{dB} \\
			\hline
		\end{tabular}
	\end{center}
\end{table}

The RTA AP is located near the left wall of the right apartment, and the positions of $M$ STAs are fixed inside the building, which we specify for each particular set of experiments.
We choose RTA STAs locations to make favorability vectors highly likely to differ greatly, which challenges the algorithm the most.
The RTA STAs generate 256-byte packets periodically with period $T_{RTA}$. 
For greater reliability, RTA STAs use MCS~0.
We use the channel model and penetration losses for the Residential Scenario described in \cite{TGaxScen}.
We list the other modeling parameters in Table~\ref{tab:parameters}.

\begin{figure}[tb]
	\captionsetup[subfigure]{aboveskip=50pt}
	\centering
	\subfloat[]{\includegraphics[width=0.49\linewidth]{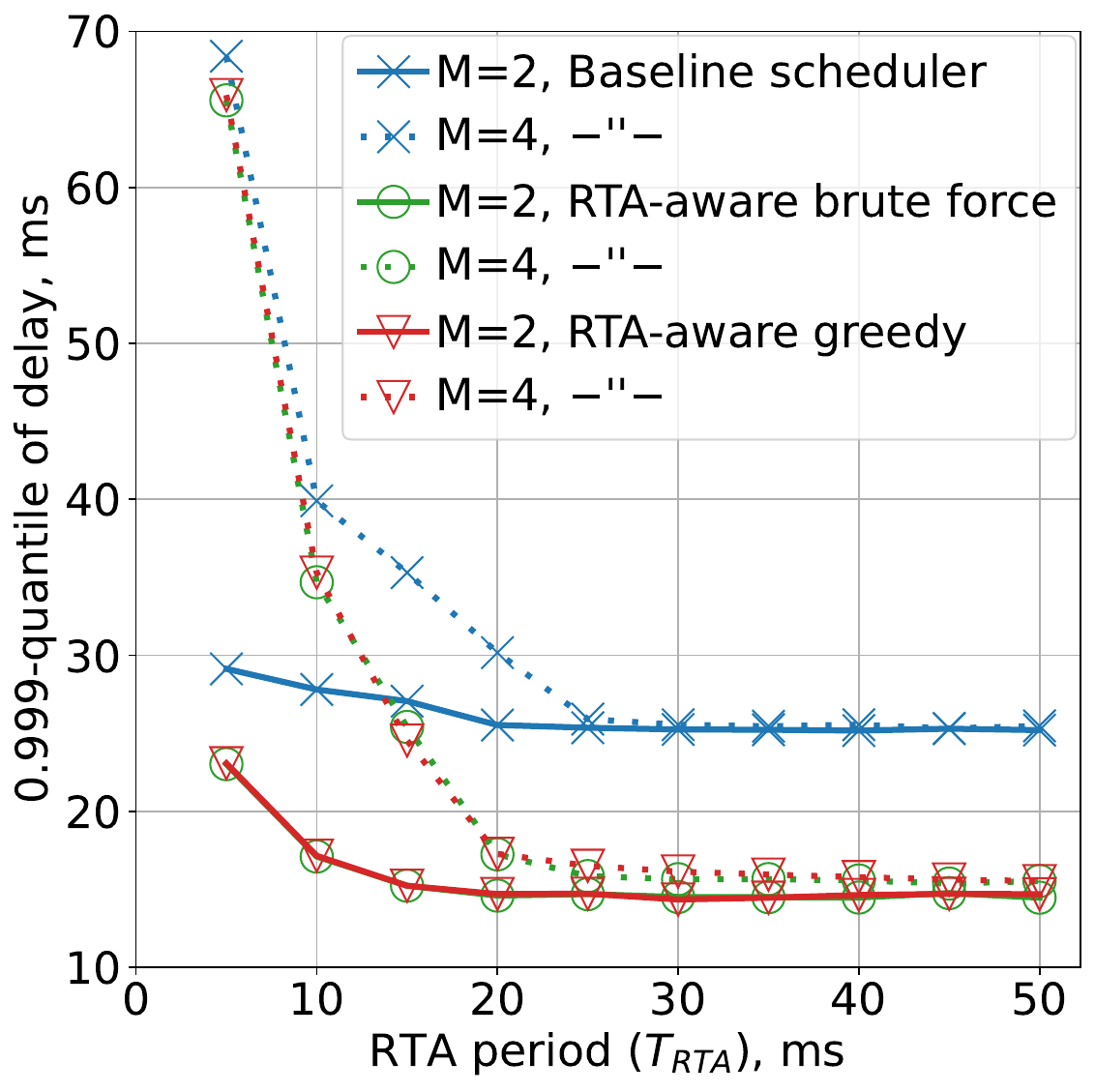}}
	\subfloat[]{\includegraphics[width=0.49\linewidth]{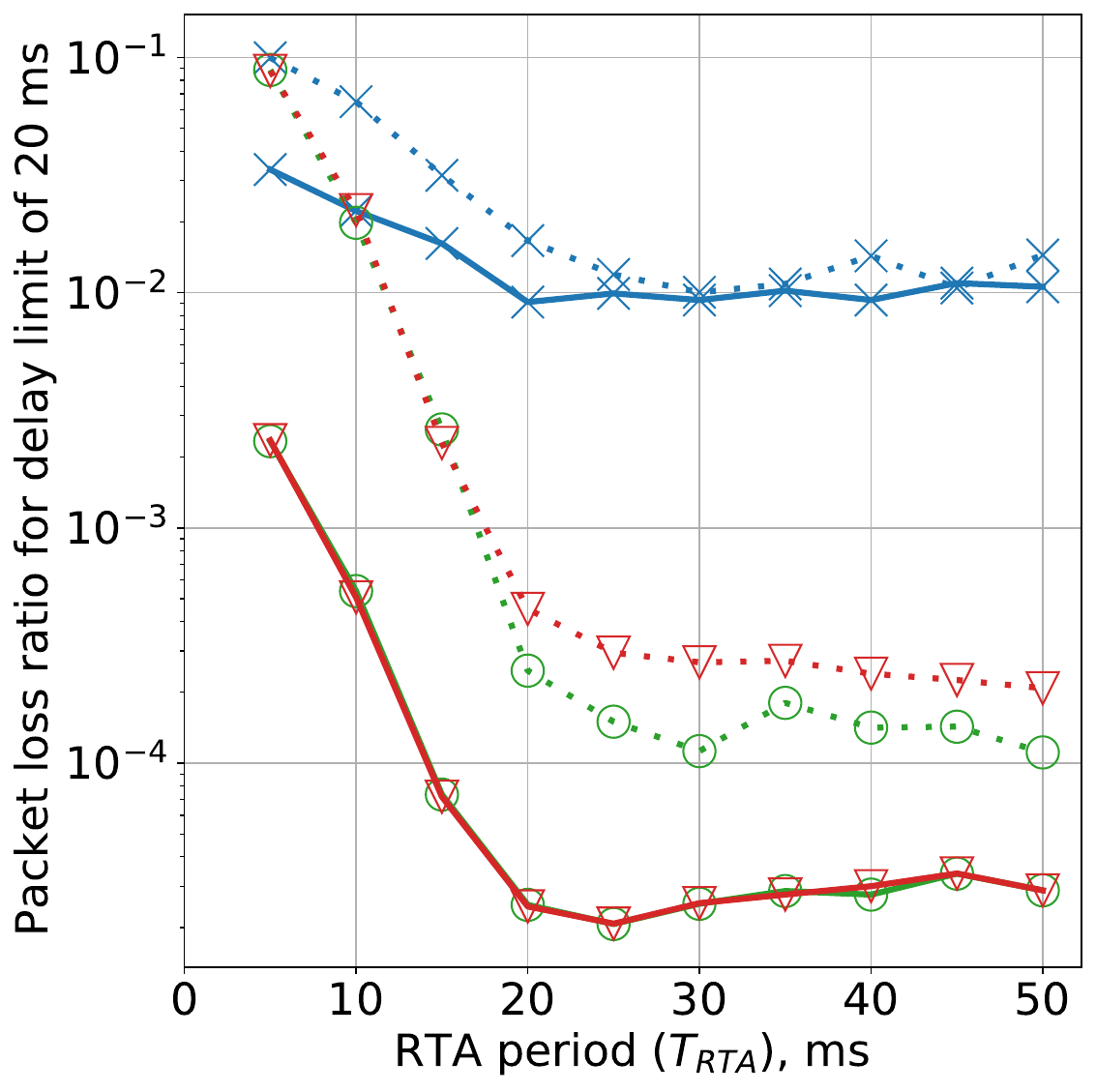}} 
	\\
	\subfloat[]{\includegraphics[width=0.49\linewidth]{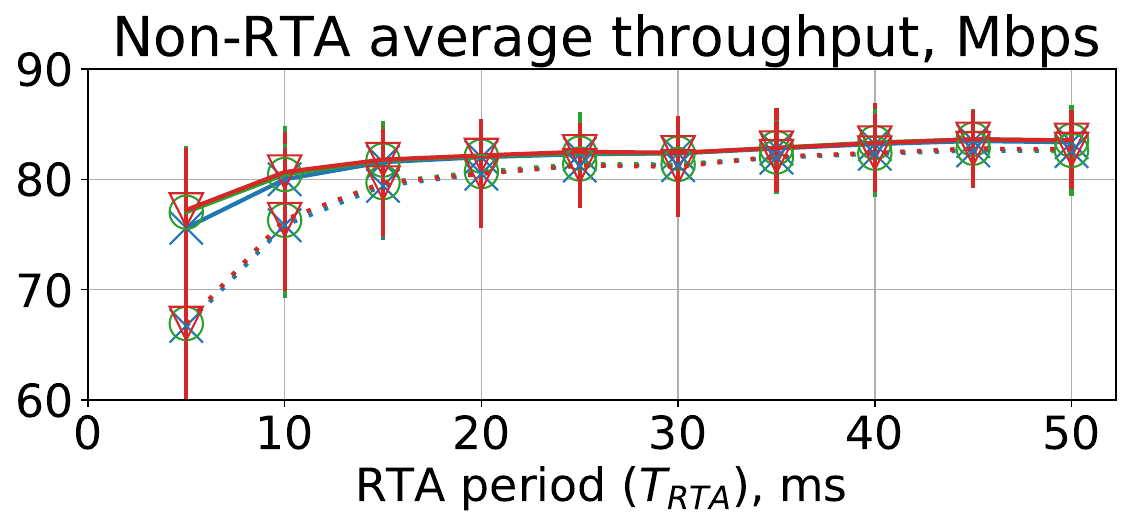}}
	\subfloat[]{\includegraphics[width=0.49\linewidth]{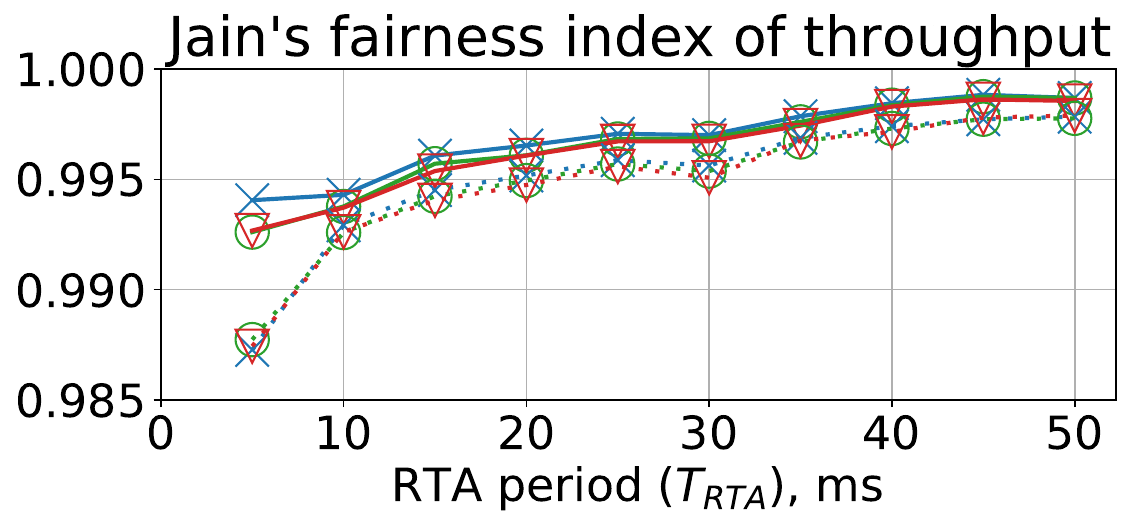}}
	\caption{\label{fig:plot} The QoS metric. (a) 0.999-quantile of delay. (b) Packet loss ratio. (c) Average throughput. (d) Jain's fairness index of throughput.}
\end{figure}

We run two sets of experiments.
For the first set of experiments, we place $M=2$ RTA STAs as shown in Fig.~\ref{fig:scenario}.
We vary $T_{RTA}$ and compare QoS metrics for RTA and non-RTA BSSs for the baseline airtime fairness scheduler and two RTA-aware schedulers: brute force and greedy solution of problem~\eqref{eq:opt_problem}.
As shown in Fig.~\ref{fig:plot} (solid lines), for $T_{RTA}>\SI{15}{ms}$, the RTA-aware schedulers almost halve the 0.999-quantile of delay.
Moreover, for the delay limit of \SI{20}{ms}, the RTA-aware schedulers reduce the packet loss ratio by two orders.

We average the performance for each arrangement over all arrangements, see Fig.~\ref{fig:plot}.
RTA-aware schedulers do not worsen the average throughput of non-RTA BSS compared to the baseline scheduler because RTA STAs use PSR opportunities more often and consume less channel time.
Jain's fairness index of throughput is close to 1 for all the studied schedulers because they guarantee equal distribution of resources.

Let us compare two RTA-aware schedulers.
According to all metrics, the results for the brute force and the greedy algorithm are almost the same.
The greedy algorithm loses in the delay quantile at most 2\%.
Thus, the results confirm that although the greedy algorithm cannot guarantee the optimal solution to the problem \eqref{eq:opt_problem}, it manages to find the best non-RTA transmission schedule in the vast majority of cases.

For the second set of experiments, we place $M=4$ RTA STAs as shown in Fig.~\ref{fig:scenario} and measure the same metrics (see Fig.~\ref{fig:plot}, dashed lines).
Despite the higher number of RTA STAs, the RTA-aware schedulers still give a 40\% gain in the quantile of delay and decrease the ratio of frames not served within \SI{20}{ms} by almost 100 times.
In this scenario, the greedy algorithm again shows results close to the brute force one.

The results show that the baseline scheduler does not allow serving RTAs, for instance, immersive communications that require 99.9\% of packets to be served with delays below \SI{20}{ms}~\cite{giordano2023wifi8}. 
At the same time, the developed scheduler handles this task without compromising the QoS for non-RTA traffic.

\section{Conclusion}
\label{sec:conclusion}

In this letter, we tune PSR with multi-AP coordination and an RTA-aware scheduler for delay-tolerant traffic to service several RTA STAs in overlapping Wi-Fi~8 networks. 
To develop the RTA-aware scheduler, we state an optimization problem to minimize the worst-case channel access delays and propose a greedy solution.
Extensive simulation shows that, with our scheduler, the average throughput and fairness index of throughput for non-RTA traffic do not change in comparison with the airtime fairness scheduler, but the 0.999-quantile of delay for RTA traffic almost halved.
Moreover, we show that our fast greedy algorithm gives a solution that differs from the optimal one by less than 2\% in terms of delay quantile.

\bibliographystyle{IEEEtran}
\bibliography{biblio}

\end{document}